\documentclass[sigconf,natbib=true,dvipsnames]{acmart}
\pdfoutput=1

\usepackage{bm}
\usepackage{graphicx}
\usepackage{float}
\usepackage[utf8]{inputenc}
\usepackage{enumitem}
\usepackage{amsfonts}
\usepackage[ruled,linesnumbered, noend]{algorithm2e}
\usepackage{pgfplots}
\usepackage{makecell}
\usepackage{arydshln}
\usepackage{subcaption}
\usepackage{tikz}
\usepackage{adjustbox}
\usepackage{booktabs}
\usepackage{threeparttable}
\usepackage{comment}
\usepackage{mathtools}
\usepackage[marginal]{footmisc}
\usepackage{balance}

\AtBeginDocument{%
  \providecommand\BibTeX{{%
    \normalfont B\kern-0.5em{\scshape i\kern-0.25em b}\kern-0.8em\TeX}}}

\copyrightyear{2022}
\acmYear{2022}
\setcopyright{acmcopyright}\acmConference[CIKM '22]{Proceedings of the 31st ACM International Conference on Information and Knowledge Management}{October 17--21, 2022}{Atlanta, GA, USA}
\acmBooktitle{Proceedings of the 31st ACM International Conference on Information and Knowledge Management (CIKM '22), October 17--21, 2022, Atlanta, GA, USA}
\acmPrice{15.00}
\acmDOI{10.1145/3511808.3557274}
\acmISBN{978-1-4503-9236-5/22/10}

\begin{document}

\title{CROLoss: Towards a Customizable Loss for Retrieval Models in Recommender Systems}




\author{Yongxiang Tang}
\affiliation{%
  \institution{Alibaba Group}
  \city{Beijing}
  \country{China}}
\email{tangyongxiang94@gmail.com}

\author{Wentao Bai}
\affiliation{%
  \institution{Alibaba Group}
  \city{Beijing}
  \country{China}}
\email{bwt1997@126.com}

\author{Guilin Li}
\affiliation{%
  \institution{Alibaba Group}
  \city{Beijing}
  \country{China}}
\email{liguilin.lgl@lazada.com}

\author{Xialong Liu}
\affiliation{%
  \institution{Alibaba Group}
  \city{Beijing}
  \country{China}}
\email{xialong.lxl@alibaba-inc.com}

\author{Yu Zhang}
\affiliation{%
  \institution{Alibaba Group}
  \city{Beijing}
  \country{China}}
\email{daoji@lazada.com}

\renewcommand{\shortauthors}{Yongxiang Tang et al.}

\begin{abstract}
In large-scale recommender systems, retrieving top N relevant candidates accurately with resource constrain is crucial. 
To evaluate the performance of such retrieval models, Recall@N, the frequency of positive samples being retrieved in the top N ranking, is widely used. 
However, most of the conventional loss functions for retrieval models such as softmax cross-entropy and pairwise comparison methods do not directly optimize Recall@N. 
Moreover, those conventional loss functions cannot be customized for the specific retrieval size N required by each application and thus may lead to sub-optimal performance. 
In this paper, we proposed the {\bfseries C}ustomizable  {\bfseries R}ecall@N {\bfseries O}ptimization {\bfseries Loss} ({\bfseries CROLoss}), a loss function that can directly optimize the Recall@N metrics and is customizable for different choices of $N$s. 
This proposed CROLoss formulation defines a more generalized loss function space, covering  most of the conventional loss functions as special cases. Furthermore, we develop the Lambda method, a gradient-based method that invites more flexibility and can further boost the system performance. We evaluate the proposed CROLoss on two public benchmark datasets. 
The results show that CROLoss  achieves  SOTA results over conventional loss functions for both datasets with various choices of retrieval size N .
CROLoss has been deployed onto our online E-commerce advertising platform, where a fourteen-day online A/B test demonstrated that  CROLoss 
contributes to a  significant business revenue growth of 4.75\%.
\renewcommand{\thefootnote}{}
\footnote{\noindent Yongxiang Tang and Wentao Bai contributed equally.}
\footnote{\noindent Code implementation is available at https://github.com/WDdeBWT/CROLoss/}
\end{abstract}


\keywords{Recommender systems, information retrieval, Recall@N optimization}
\maketitle

\section{Introduction}
Large-scale recommender systems are widely adopted in commercial applications, such as media recommendation \cite{youtubeDNN, google2tower}, commodities recommendation \cite{AliComirec,EGES} and computational advertising \cite{TDM,DIN}. With the scale of users and items on such popular platforms expanding rapidly, multi-stage recommender systems \cite{youtubeDNN,google2tower,fbEmbedding} have become an important strategy. The multi-stage recommender system is consist of the retrieval stage \cite{TDM, houle2014rank, muja2014scalable} that selects hundreds of candidates from the whole item set and the ranking stage \cite{wideanddeep, guo2017deepfm} that sorts the candidates in a more precise order before final exposure.  In this paper, we focus on optimizing the retrieval stage, of which the goal is to optimize the Recall@N metrics, which is defined as the frequency of positive samples being retrieved in the top N ranking. 

Popular retrieval models such as \cite{google2tower,AliComirec,  TDM,DSSM, youtubeDNN, fbEmbedding, EGES, PinSAGE, Octopus, LightGCN} mainly adopt softmax cross-entropy or pairwise comparison methods \cite{triplet,rendle2012bpr} as their loss functions. The softmax cross-entropy method 
transforms the retrieval scores into a probability distribution and minimizes the cross-entropy loss. The pairwise comparison methods including the triplet loss \cite{triplet} and some learning to rank methods such as RankNet \cite{RankNet} and BPR loss \cite{rendle2012bpr} optimize the retrieval model by learning the order between positive and negative items for each user. 
Although these methods have achieved great success in practical recommender systems, neither the the cross-entropy loss or the pairwise comparison methods is directly optimizing the Recall@N metrics. Moreover, as well known, a proper choice of retrieval size $N$ provides a trade-off between system efficiency and accuracy for different commercial usage. However, these conventional losses  are also not adaptable for different choice of retrieval size $N$, which may lead to sub-optimal solutions for practical commercial usage. 

In this paper, we propose the {\bfseries C}ustomizable {\bfseries R}ecall@N {\bfseries O}ptimization {\bfseries Loss} ({\bfseries CROLoss}), 
which is derived to directly optimize the Recall@N metrics. 
We first rewrite the Recall@N metrics in the form of pairwise sample comparison. By leveraging a pairwise comparison kernel function $\phi$, this objective function is derived as a differentiable loss functional space.
We also invite a  weighting function $w_{\alpha}$  to allow this loss function customizable for a different choice of $N$. Besides, it can be shown that the proposed CROLoss function space covers the conventional cross-entropy loss, triplet loss, and BPR loss as special cases. Furthermore, by analyzing the gradient of the CROLoss, we find that the comparison kernel $\phi$ plays two different roles. To further improve this loss function, we develop the Lambda method, a gradient-based method that allows a different choice of kernel $\phi_1$ and $\phi_2$ for these two roles, and further boost the system performance.

Extensive experiments are conducted on two public benchmark datasets. Experimental results demonstrate that the proposed CROLoss can significantly improve traditional loss functions on both datasets for all different choices of N. These experiments also verify the customizability of CROLoss for different retrieval sizes $N$ through adjusting the comparison kernels and weighting parameters. Furthermore, a fourteen-day online A/B is performed on our online E-commerce advertising platform, where CROLoss improved Recall@N by 6.50\%  over the cross-entropy loss and contributes to a  significant business revenue growth of 4.75\%.   

The main contributions of our paper are summarized as follows:
\begin{itemize}
    \item We propose the CROLoss, to our best knowledge, it is the first method that directly optimizes the Recall@N metrics and can be customized for different retrieval sizes $N$.
    \item This CROLoss covers a more generalized loss function space, allowing users to tailor it for their specific usage by tuning the kernel and the weighting parameter. We proved that cross-entropy loss, triplet loss and BPR loss are special cases of our method.
    \item We develop the Lambda method, a gradient-based method that allows a more flexible choice of the kernel function and further boost the system performance. 
    \item We conduct extensive experiments on two real-world public datasets, which show that CROLoss outperforms existing loss functions significantly.
\end{itemize}

\section{Related Work}
The retrieval stage  aims to retrieve $N$ items from a large corpus with both the efficiency and the accuracy as system objective  \cite{youtubeDNN}. In recent years, extensive work has been developed to improve the retrieval model by  learning high-quality user and item representation vectors \cite{AliMIND,AliComirec,Octopus,TDM,EGES,PinSAGE} or by organizing the  training data in a wise manner \cite{google2tower,ma2020disentangled}. Limited attention has been paid to the design of a more suitable loss function for this task. Most conventional models employ two categories of retrieval loss: the softmax cross-entropy loss and the pairwise method. However, all these losses do not directly optimize  Recall@N which may lead to suboptimal solutions. In our practice, we also found that a fixed loss form may not be suitable for every different choice of $N$. 
\subsection{Retrieval Model Optimization Losses}
The softmax cross-entropy loss treats the optimization of the retrieval model as a classification problem \cite{google2tower, AliComirec, TDM, DSSM}.  Suppose item $v$ is a positive sample for user $u$ taken from a set of positive user-item pairs $\mathcal{C}$, and items $v_1,\cdots,v_n$ are negative samples for user $u$. The softmax function transforms the retrieval scores $S(u,v_i)$ to a probability distribution, and we have the cross-entropy minimization objective:
\begin{equation}
	\mathcal{L}_{softmax} = -\sum_{(u,v) \in \mathcal{C}} \log\frac{e^{S(u,v)}}{e^{S(u,v)}+\sum_i e^{S(u,v_i)}},
    \label{sfx}
\end{equation}
of which solution is the maximum likelihood estimation of the classification problem.

The second category is the pairwise methods \cite{fbEmbedding,PinSAGE}, which model the order between positive and negative items for each user by pairwise comparison on retrieval scores. For example, when using the user as the anchor, the triplet loss \cite{triplet} is:
\begin{equation}
	\mathcal{L}_{triplet} = \sum_{(u,v) \in \mathcal{C}} \sum_i(S(u,v_i)-S(u,v)+m)_+,
    \label{tri}
\end{equation}
where $(\cdot)_+$ is the positive part function. Since the margin parameter $m$ appears in the loss, the method can enlarge the gap between the positive and negative retrieval scores. However, both \eqref{sfx} and \eqref{tri} do not directly optimize Recall@N, and cannot be customized for the retrieval size $N$.

\subsection{Learning to Rank}

Learning To Rank (LTR) is a method for directly learning the ordering of samples, which can be classified into three categories: pointwise method, pairwise method, and listwise method.  In Section 3, we show that our CROLoss can also be viewed as a pairwise LTR method.

The pointwise methods\cite{cossock2008statistical,li2007mcrank} are not an optimal choice for retrieval task since the retrieval stage generally  does not  need to accurately estimate the  ranking score.  The listwise methods\cite{burges2010rankoverview,cao2007learning,duan2010empirical,taylor2008softrank,xia2008listwise,xu2007adarank} are generally used in scenarios where the correlation of each ranking position can be accurately determined, which is also not completely consistent with the object of the retrieval stage. The pairwise methods\cite{RankNet,rendle2012bpr,xgboost,PRMrank,LambdaRank,PolyRank} approximate the LTR problem to the classification problem, which produce ranking results by learning the relative order of each sample pair. The triplet loss mentioned above can also be seen as a form of the pairwise LTR. RankNet loss \cite{RankNet} is the classic form of pairwise LTR loss. If $S$ is a set of sample pairs, including the rank order label between samples $a$ and $b$, we have:
\begin{equation}
	\mathcal{L}_{ranknet} = -\sum_{a,b \in S} t_{ab}\log(\sigma(s_a-s_b)) + (1-t_{ab})\log(1-\sigma(s_a-s_b)),
    \label{rknet}
\end{equation}
where $s_a$ and $s_b$ are the scores predicted by the ranking model, $\sigma$ is logistic sigmoid function, $t_{ab}$ is equal to 1 if sample $a$ is ranked higher than $b$, and equal to 0 otherwise. If samples $a$ and $b$ have the same rank, then $t_{ab}$ is equal to 0.5. In the retrieval stage of the recommender system, due to the large number of candidate items, it generally uses the item clicked by the user and the item of random negative sampling to construct a sample pair, therefore, it is not necessary to consider pairs of equal order. If we manually specify that the sample $v$ is the positive sample and the sample $v_i$ is the negative sample, we can get the BPR loss \cite{rendle2012bpr}:
\begin{equation}
	\mathcal{L}_{bpr} = -\sum_{(u,v) \in \mathcal{C}} \sum_i \log(\sigma(S(u,v)-S(u,v_i))),
    \label{bpr}
\end{equation}
which is an LTR method widely used in recommender systems \cite{NGCF,LightGCN}. There are some pairwise LTR methods that use specific models \cite{xgboost,PRMrank} or require special supervision signals \cite{LambdaRank,PolyRank} and are not suitable for the retrieval stage of recommender systems.

In this paper, we will prove that $\mathcal{L}_{bpr}$, as well as the aforementioned $\mathcal{L}_{softmax}$ and $\mathcal{L}_{triplet}$, are special cases of our CROLoss, and will provide a new perspective to interpret them from the object of Recall@N optimization.

\section{Customizable Recall@N Optimization Loss}

In this section, we propose the Customizable Recall@N Optimization Loss, abbreviated as CROLoss and marked as $\mathcal{L}$. We first formulate the Recall@N optimization problem and rewrite it in the form of pairwise sample comparison using a non-differentiable comparison function $\psi$. Next, we obtain the $\mathcal{L}_{\alpha,\psi}$ by introducing a weighting function $w_\alpha(N)$ to customize the retrieval size $N$. Then, with help from the pairwise comparison kernel $\phi$, we get $\mathcal{L}_{\alpha,\phi}$, which makes CROLoss differentiable. Furthermore, we develop the Lambda method, which is an improvement on the gradient form of CROLoss and can better optimize our objective.
\subsection{The base model}
 
\begin{figure}[h]
  \centering
  \includegraphics[width=0.9\linewidth]{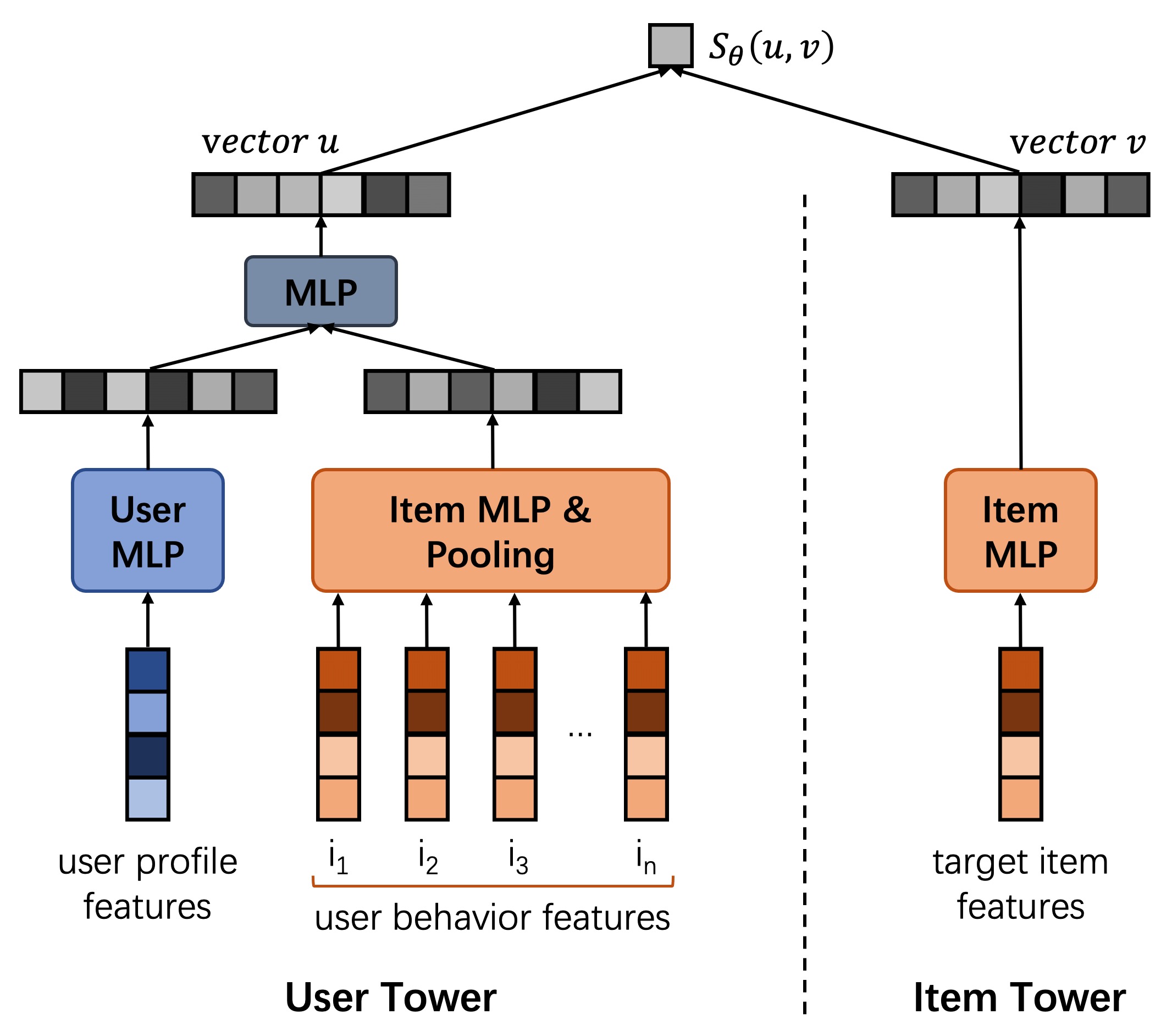}
  \caption{Architecture of two tower retrieval model}
  \label{fig:basemodel}
\end{figure}

Our approach is model-agnostic. To illustrate the idea, we adopt the widely used two-tower retrieval model \cite{youtubeDNN} as our base model, which has a good balance between accuracy and online processing speed. The model structure is shown Figure \ref{fig:basemodel}. The two-tower model can be generally divided into two parts: User Tower and Item Tower. The User Tower processes the user profile features and user behavior features to produce the user representation vector $u$, and the Item Tower processes the target item features to produce the item representation vector $v$. Finally, the retrieval score $S_{\theta}(u,v)$ of the user and the target item is generally obtained by inner product or vector similarity calculation.
 
\subsection{Problem Formulation}

Suppose the whole set of items is $\mathcal{I}=\{v_1, ..., v_{|\mathcal{I}|}\}$ and the corpus of our retrieval model is $\mathcal{C}=\{(u, v)^{(j)}\}_{j=1}^{|\mathcal{C}|}$, which is a set of positive user-item pairs. The retrieval model $S_{\theta}(u, v)$ is a scoring function for any user-item pair, and $\theta$ is its parameter. The Recall@N statistic determined by model $S_{\theta}(u, v)$ and corpus $\mathcal{C}$ is 
\begin{equation}
    {\rm Recall@N}=\frac{ | \left\{ (u,v)|S_{\theta}(u,v)\in {\rm Top_N} \{S_{\theta}(u, v_i)|v_i \in \mathcal{I}\},(u,v)\in \mathcal{C}\right\} | }{ | \mathcal{C} |},
    \label{raw_recall}
\end{equation}
where ${\rm Top_N}(\cdot)$ returns a set's largest $N$ elements. Our purpose is to find a retrieval model $S_{\theta}$ which maximizes \eqref{raw_recall}.

In order to rewrite the Recall@N metrics in a simpler form, we define the ranking of $S_{\theta}(u,v)$ among the whole user-item scoring set $\{S_{\theta}(u, v_i)|v_i \in \mathcal{I}\}$ as the ranking statistic of $(u, v)$:
\begin{equation}
    R_\psi(u, v; \theta)=
    1+\sum_{\substack{v_i \in \mathcal{I} \\ v_i \neq v}} 
    \psi \left( S_{\theta}(u, v_i)-S_{\theta}(u,v) \right),
    \label{raw_rank}
\end{equation}
where $\psi$ is the unit step function, $\psi(x) = 0$ when $x<0$ and $\psi(x)=1$ when $x>=0$. Then item $v$ is retrievable by top-N retrieval scores of user $u$ if and only if $R_\psi(u, v; \theta) \leq N$, and we can rewrite Recall@N as
\begin{equation}
{\rm Recall@N}=\frac{1}{|\mathcal{C}|} \sum_{(u,v) \in \mathcal{C}} \mathbf{1}_{\left\{ R_\psi(u, v; \theta) \leq N \right\}}\ , 1<=N<=|\mathcal{I}|.
\label{recall}
\end{equation}

\subsection{Customizable Retrieval Range}
According to the target of our retrieval model, we should pay attention to some specific ranges of $N$ when we try to optimize the Recall@N metrics. We can achieve the requirement by introducing a probability density function
\begin{equation}
    w\colon [1, |\mathcal{I}|+1) \to [0, +\infty),
\end{equation}
which satisfies
\begin{equation}
    \int_{1}^{|\mathcal{I}|+1}w(x)\, dx=1.
\end{equation}
We call $w$ the weighting function, and the value of $w(x)$ resembles the importance of Recall@x metric, where ${\rm Recall@x =Recall@\lfloor x \rfloor}$ is the generalized form of Recall@N for $x \in \mathbb{R}$.

By using $w(x)$ to weight the importance of Recall@x metrics for $x \in [1, |\mathcal{I}|+1)$, we can optimize the Recall@N metrics for all possible Ns by defining our loss function as the negative expectation of Recall@x with respect to the PDF $w(x)$ as follows,
\begin{equation}
\begin{aligned}
\mathcal{L}(\theta)
    =& 
    -\mathbb{E}_w \, {\rm Recall@x}
    \\ =& 
    -\int_{1}^{|\mathcal{I}|+1}w(x) {\rm Recall@x}\, dx
\end{aligned}
\label{crol_defination}
\end{equation}

By \eqref{recall}, we can simplify the expression of $\mathcal{L}(\theta)$ as:
\begin{equation}
\begin{aligned}
\mathcal{L}(\theta)
    =& 
    -\int_{1}^{|\mathcal{I}|+1}w(x) \sum_{(u,v) \in \mathcal{C}} \frac{1}{|\mathcal{C}|} \mathbf{1}_{\left\{ R_\psi(u, v; \theta) \leq x \right\}}\, dx,
    \\ =&
    \frac{-1}{|\mathcal{C}|}\sum_{(u,v) \in \mathcal{C}} \int_{1}^{|\mathcal{I}|+1}w(x) \mathbf{1}_{\left\{ R_\psi(u, v; \theta) \leq x \right\}}\, dx
    \\ =&
    \frac{-1}{|\mathcal{C}|}\sum_{(u,v) \in \mathcal{C}} \int_{R_\psi(u, v; \theta)}^{|\mathcal{I}|+1}w(x)\, dx 
    \\ =&
    \frac{-1}{|\mathcal{C}|}\sum_{(u,v) \in \mathcal{C}} \left( 1- \int_{1}^{R_\psi(u, v; \theta)}w(x) \right)\, dx 
    \\=&
    \frac{1}{|\mathcal{C}|}\sum_{(u,v) \in \mathcal{C}} W(R_\psi(u, v; \theta)) - 1,
\end{aligned}
\label{crol_derive}
\end{equation}
where $W(x)=\int_1^x w(z)\,dz$ is the cumulative distribution function of $w(x)$.

We generally hope the value of $w(x)$ is larger for smaller $x$ since for most recommender systems, the retrieval size is far less than the size of the candidate set $|\mathcal{I}|$, making most of the Recall@N metrics with greater $N$ unimportant. Here, we take the power function $w_\alpha(x)$ as our weighting function:
\begin{equation}
    w_{\alpha}(x)=\frac{1}{Z}x^{-\alpha}
    \label{W}
\end{equation}
where $\alpha \leq 0$ is the weighting parameter and $Z$ is a regularization parameter so that $\int_{1}^{|\mathcal{I}|+1}w_{\alpha}(x)\, dx=1$. Then $W(N)$ becomes
\begin{equation}
    W_{\alpha}(N)=\int_{1}^{N} w_{\alpha}(x)\, dx=
    \begin{dcases}
    \frac{\log(N)}{\log(|\mathcal{I}|+1)}  & \text{if $\alpha=1$}, \\
    \frac{1-N^{1-\alpha}}{1-(|\mathcal{I}|+1)^{1-\alpha}} & \text{else.}
    \end{dcases}
\end{equation}
Therefore, by neglecting the constant $-1$ in \eqref{crol_derive}, we have the loss function controlled by $\alpha$:
\begin{equation}
    \mathcal{L}_{\alpha,\psi}(\theta) = \sum_{(u,v) \in \mathcal{C}} W_{\alpha}(R_\psi(u, v; \theta)).
    \label{crol}
\end{equation}
For a larger $\alpha$, the weighting function $w_{\alpha}(x)$ decays faster and guides our model to optimize Recall@N on smaller $N$s. Hence, we can customize the Recall@N optimization problem by tuning the weighting parameter $\alpha$.

\subsection{Pairwise Kernel-Based Objective}

We notice that our objective in \eqref{crol} is not differentiable, which means it is impossible to solve the problem via a gradient descent algorithm. To solve this issue, we modify the unit step function $\psi$ into a differentiable substitution $\phi(x)$, then we have the approximation of the ranking statistic in \eqref{raw_rank}:
\begin{equation}R_{\phi}(u, v; \theta)=
    1+\sum_{\substack{v_i \in \mathcal{I} \\ v_i \neq v}} \phi(S_{\theta}(u, v_i)-S_{\theta}(u,v)) .
    \label{rank_phi}
\end{equation}

By \eqref{rank_phi}, we have the approximation of $\mathcal{L}_{\alpha,\psi}(\theta)$ in \eqref{crol}:
\begin{equation}
\begin{aligned}
\mathcal{L}_{\alpha,\phi}(\theta)=& 
    \sum_{(u,v) \in \mathcal{C}} W_{\alpha} \left(R_{\phi}(u, v; \theta) \right)\\
    =& \sum_{(u,v) \in \mathcal{C}} W_{\alpha} \left(1+\sum_{\substack{v_i \in \mathcal{I} \\ v_i \neq v}} \phi(S_{\theta}(u, v_i)-S_{\theta}(u,v)) \right),
\end{aligned}
\label{crol_alpha_phi}
\end{equation}
which is the formal form of the CROLoss.

We call function $\phi(x)$ the comparison kernel if it satisfies the following conditions: 

i) $\phi(x)$ is differentiable almost everywhere; 

ii)  $\phi(x)$ is monotonically increasing; 

iii) $\lim_{x \to -\infty} \phi(x) = 0$; 

iv) $0.5<=\phi(0)<=1$; 

v) $\lim_{x ->\infty} \phi(x) >= 1$.

\begin{figure}[t]  
\centering
\begin{tikzpicture}[scale=0.9]
\begin{axis}[
    legend style={at={(0.03,0.97)},
    anchor=north west},
    legend cell align={left},
    font=\Large
]
\addplot [
    thick,
    domain=-2:2, 
    samples=21, 
    color=black,
    line width=1.3pt,
    style=dashed
    ] coordinates {
        (-2.0,0)
        (0,0)
        (0,1)
        (2.0,1)
        }
    ;
\addplot [
    thick,
    domain=-2:2, 
    samples=21, 
    color=violet,
    line width=1.3pt,
    ]
    {1 / (1+e^(-x))}
    ;
\addplot [
    thick,
    domain=-2:1.4, 
    samples=18, 
    color=olive,
    line width=1.3pt,
    ]
    {e^x}
    ;
\addplot [
    thick,
    domain=-2:2, 
    samples=21, 
    color=teal,
    line width=1.3pt,
    ]
    {0.5 * abs(x+0.8) * (1+sign(x+0.8))}
    ;
\addplot [
    thick,
    domain=-2:2, 
    samples=21, 
    color=orange,
    line width=1.3pt,
    ]
    {ln(1+e^(x))}
    ;
\legend{Unit step, Sigmoid, Exponential, Hinge, Softplus};
\end{axis}
\end{tikzpicture}
\caption{Example of optional comparison kernel functions.}
\label{fig_kernel}
\end{figure}
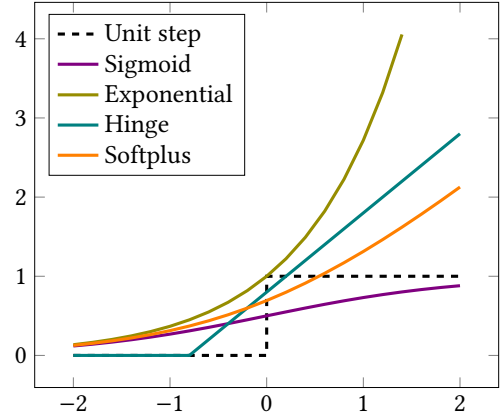

With these properties, $\phi$ can approximates $\psi$ well for $x < 0$. However, we do not strictly constraint $\phi(x) \leq 1$ for $x \geq 0$, since the optimization can be easier when $\phi(x)$ is convex. For a convex kernel $\phi$, larger gaps in $G_u=\{S_{\theta}(u,v_i)-S_{\theta}(u,v)|v_i \in \mathcal{I}, v_i \neq v\}$ can decay faster via SGD \cite{sgd} due to the convexity of $\phi$, letting positive gap $g \in G_u$ close to 0 and $\phi(g)$ close to $\phi(0)$, where $\phi(0) \in [0.5, 1]$. Thus, it prevents \eqref{crol_alpha_phi} go too far from the original $\mathcal{L}_{\alpha,\psi}$. As shown in Figure \ref{fig_kernel}, there are many popular activation functions can be regarded as the comparison kernel. We will verify the adaptability of these kernels to CROLoss in the experimental section.


\subsection{Relation to Existing Methods}

In this section, we will prove that by choosing specific comparison kernel and alpha, the softmax cross-entropy loss $\mathcal{L}_{softmax}$, the triplet loss $\mathcal{L}_{triplet}$ and the BPR loss $\mathcal{L}_{bpr}$ can be regarded as special cases of our CROLoss. 

By taking $\phi(x)=\mathrm{exp}(x)=e^x$ and $\alpha=1$, we have
\begin{equation}
\begin{aligned}
\mathcal{L}_{1,\mathrm{exp}}(\theta)
=& \sum_{(u,v) \in \mathcal{C}} \log \left(1 + \sum_{\substack{v_i \in \mathcal{I} \\ v_i \neq v}} e^{S_{\theta}(u, v_i)-S_{\theta}(u,v)} \right)\\
=& \sum_{(u,v) \in \mathcal{C}} \log \left(\frac{\sum_{v_i \in \mathcal{I}} e^{S_{\theta}(u, v_i)}}{e^{S_{\theta}(u,v)}} \right)\\
=& -\sum_{(u,v) \in \mathcal{C}} \log \left(\frac{e^{S_{\theta}(u,v)}}{\sum_{v_i \in \mathcal{I}} e^{S_{\theta}(u, v_i)}} \right),
\end{aligned}
\label{crol_sfx}
\end{equation}
which is identical to the softmax cross-entropy loss shown in \eqref{sfx}.

By taking $\phi(x)=\mathrm{hinge}(x)=(x+m)_+$ and $\alpha=0$, we have
\begin{equation}
\begin{aligned}
\mathcal{L}_{0,\mathrm{hinge}}(\theta)
=&
\frac{1}{|\mathcal{I}|} \left(1+ \sum_{(u,v) \in \mathcal{C}} \, \sum_{\substack{v_i \in \mathcal{I} \\ v_i \neq v}} \left(S_{\theta}(u, v_i)-S_{\theta}(u,v)+m \right)_+  \right) \\
=&
\frac{1}{|\mathcal{I}|} \sum_{(u,v) \in \mathcal{C}} \, \sum_{\substack{v_i \in \mathcal{I} \\ v_i \neq v}} \left(S_{\theta}(u, v_i)-S_{\theta}(u,v)+m \right)_+ + const,
\label{crol_tri}
\end{aligned}
\end{equation}
which is identical to the triplet loss shown in \eqref{tri}.

By taking $\phi(x)=\mathrm{softplus}(x)=\log(1+e^x)$ and $\alpha=0$, we have
\begin{equation}
\begin{aligned}
\mathcal{L}_{0,\mathrm{softplus}}(\theta)
=& \frac{1}{|\mathcal{I}|} \left( 1 + \sum_{(u,v) \in \mathcal{C}} \sum_{\substack{v_i \in \mathcal{I} \\ v_i \neq v}} \log \left(1 + e^{S_{\theta}(u, v_i)-S_{\theta}(u,v)} \right) \right) \\
=& \frac{1}{|\mathcal{I}|}\sum_{(u,v) \in \mathcal{C}} \sum_{\substack{v_i \in \mathcal{I} \\ v_i \neq v}} -\log \left(\frac{1}{1 + e^{-\left(S_{\theta}(u, v)-S_{\theta}(u,v_i)\right)}} \right) + const,
\end{aligned}
\label{crol_bpr}
\end{equation}
which is identical to the BPR loss shown in \eqref{bpr}.

Therefore, we provide a new perspective to interpret these conventional methods in the form of CROLoss. We will further analyze the impact of their selection of specific kernels and alphas on performance in the experiments.

\subsection{The Lambda Method}

Introducing the comparison kernel is helpful for optimizing the CROLoss. However, the range of most comparison kernels is $(0, \infty)$, causing the prediction of the ranking statistic $R_\phi(u, v; \theta)$ imprecise and thus inability to accurately estimate the gradient of the parameters through the weight function. To find an appropriate method to optimize the model parameters, we focus on the gradient of CROLoss to the parameters. We let $g_u^{(i)}=S_{\theta}(u, v_i)-S_{\theta}(u,v)$ to represent the gap between negative score $S_{\theta}(u, v_i)$ and positive score $S_{\theta}(u,v)$. Then, according to \eqref{crol_alpha_phi} we have the partial derivative of $\mathcal{L}_{\alpha,\phi}(\theta)$ with respect to $g_u^{(i)}$:

\begin{equation}
\frac{\partial \mathcal{L}_{\alpha,\phi}(\theta)}{\partial g_u^{(i)}} =
 \left\{ w_{\alpha}(R_{\phi}(u, v; \theta)) \right\} \left\{ \phi' (g_u^{(i)}) \right\}.
\label{crol_gradient}
\end{equation}

We use curly brackets to divide the right-hand side of \eqref{crol_gradient} into two parts, the first part can be regarded as the approximated weighting density, and the second part can be regarded as the descendent velocity of each gap $g_u^{(i)}$. The choice of the comparison kernel function $\phi$ is the key to the performance of CROLoss. For the first part, we want to estimate $R_{\phi}(u, v; \theta)$ accurately and map it to an appropriate weighting density.  As shown in Figure \ref{fig_kernel}, taking $\phi$ as the logistic sigmoid or the unit step function can bring $R_{\phi}(u, v; \theta)$ closer to the authentic ranking statistic $R_{\psi}(u, v; \theta)$. For the second part, we hope that $\phi$ can properly penalize on larger $g_u^{(i)}$ to speed up model fitting, so the choice of $\phi$ can be softplus function, hinge function, or exponential function. Therefore, the comparison kernel $\phi$ actually plays two roles, and their goals are inconsistent. Inspired by \cite{LambdaRank}, which directly converts the non-differentiable metric into a multiplier on the gradient of the objective function, we propose the Lambda method to resolve this conflict. 

First, we use a multiplier $\lambda^{(u)}$ to replace $w_{\alpha}(R_{\phi}(u, v; \theta))$ in \eqref{crol_gradient}, assuring $\lambda^{(u)}$ to be a suitable estimation of the actual weighting density $w_{\alpha}(R_{\psi}(u, v; \theta))$. We take

\begin{equation}
    \lambda_{\alpha,\phi_1}^{(u)} = w_{\alpha} \left( R_{\phi_1}(u, v; \theta) \right)
    = w_{\alpha} \left( 1+\sum_{\substack{v_i \in \mathcal{I} \\ v_i \neq v}} \phi_1(S_{\theta}(u, v_i)-S_{\theta}(u,v)) \right),
\end{equation}
which is regarded as a non-differentiable value during training.

Then, we take a suitable kernel $\phi_2$ to replace the kernel  $\phi$ of $\phi' (g_u^{(i)})$ in \eqref{crol_gradient}, and the partial derivative form shown in \eqref{crol_gradient} can be replaced by 

\begin{equation}
    \frac{\partial \mathcal{L}_{\alpha,\phi_1, \phi_2}(\theta)}{\partial g_u^{(i)}} =\lambda_{\alpha,\phi_1}^{(u)} \phi_2' (g_u^{(i)}),
\label{lambda_partial}
\end{equation}
where $\mathcal{L}_{\alpha,\phi_1, \phi_2}$ is the new loss function controlled by two independent comparison kernel $\phi_1$ and $\phi_2$.

Therefore, we can configure different choices for $\phi_1$ and $\phi_2$, and even choose a non-differentiable function such as a unit step function directly in $\phi_1$. Based on the form of partial derivative controlled by $\phi_1$ and $\phi_2$ in \eqref{lambda_partial}, we can derive the original loss function of the lambda method:

\begin{equation}
\begin{aligned}
\mathcal{L}_{\alpha,\phi_1, \phi_2}(\theta)
=& \sum_{(u,v)\in \mathcal{C}} \lambda_{\alpha,\phi_1}^{(u)} \left( 1+\sum_{\substack{v_i \in \mathcal{I} \\ v_i \neq v}} \phi_2 \left( g_u^{(i)} \right) \right)\\
=& \sum_{(u,v)\in \mathcal{C}} {\rm SG} \left( w_{\alpha} \left( R_{\phi_1}(u, v; \theta) \right) \right)  R_{\phi_2}(u, v; \theta) ,
\end{aligned}
\label{crol_lambda}
\end{equation}
where ${\rm SG}$ resembles the stop-gradient method, that avoids the gradient backpropagation from $\lambda_{\alpha,\phi_1}^{(u)}$ to the model parameter $\theta$.

\section{Experiment}
In this section, we conduct extensive experiments on two benchmark datasets, as well as an online A/B test in our advertising system, to answer the following questions: 

{\bfseries RQ1}: Could we boost the performance of retrieval models with different retrieval size $N$s by using CROLoss?

{\bfseries RQ2}: How to select the comparison kernel $\phi$ and weighting parameter $\alpha$ for different retrieval size $N$s?

{\bfseries RQ3}: Whether the Lambda method can further improve the performance of CROLoss, and how to choose suitable kernel functions for $\phi_1$ and $\phi_2$?

{\bfseries RQ4}: Can CROLoss improve the performance of current losses in a live recommender system?

\subsection{Experimental Setup}

\begin{table}[h]
  \caption{Dataset basic statistics}
  \label{dataset_statistics}
  \begin{tabular}{lccc}
    \toprule
    Dataset & \#users & \#items & \#behaviors\\
    \midrule
    Amazon Books & 459,133 & 313,966 & 8,898,041\\
    Taobao & 976,779 & 1,708,530 & 85,384,110\\
    \bottomrule
  \end{tabular}
\end{table}

\noindent {\bfseries Datasets.} We conduct experiments on two public real-world datasets. The statistics of the two datasets are shown in Table \ref{dataset_statistics}.

\begin{itemize}
    \item {\bfseries Amazon Books} \cite{books} is the \emph{Books} category of the Amazon dataset, which is consist of product reviews from Amazon.
    \item {\bfseries Taobao} \cite{TDM} is a public dataset collected users' click logs from Taobao’s recommender systems.
\end{itemize}

We split all users into training/validation/test sets by the proportion of 8:1:1 and use the user's first $k$ behaviors to predict the $(k+1)$-th behavior in each training sample. We truncate the length of user behavior sequences to 20 for the Amazon Book dataset and 50 for the Taobao dataset.

\noindent {\bfseries Compared Methods.} To evaluate the performance of our proposed method, we compare the CROLoss with the following conventional methods:

i) softmax cross-entropy loss shown in \eqref{sfx};

ii) triplet loss shown in \eqref{tri};

iii) BPR loss shown in \eqref{bpr}.

Besides, we test the performance of CROLoss under different comparison kernels and weighting parameters to reveal their effects. For the Lambda method, we regard it as a version of CROLoss and include it in the comparison. The optional kernel functions are as follows:

i) hinge kernel $\phi_{hinge}(x) = (x+m)_+$;

ii) logistic sigmoid kernel $\phi_{sigmoid}(x) = 1/(1+e^{-x})$;

iii) exponential kernel $\phi_{exponential}(x) = e^x$;

iv) softplus kernel $\phi_{softplus}(x) = \log(1+e^x)$.





\noindent {\bfseries Implementation Details.} 
 We take $S(u, v)=\tau\langle u, v\rangle$, where $\langle u, v\rangle$ is the cosine similarity of user vector $u$ and item vector $v$. $\tau$ is the scale parameter to map the $S(u, v)$ to the appropriate range of the kernel function, we generally set it to 10 to achieve the best performance.

To speed up model training, we uniformly sample $n_{rn}$ random items for each positive sample and share these random items inside a mini-batch of size $n_{bs}$, letting each positive sample be attached with a set of $n_{rn}*n_{bs}$ random items acting as negative items. For uniformly sampled subset $\mathcal{I}'$ of $\mathcal{I}$, a biased but useful estimation of the original ranking statistic in \eqref{raw_rank} is
\begin{equation}
    \hat{R}_{\phi}(u, v; \theta)=\frac{|\mathcal{I}|}{|\mathcal{I}'|} \left\{1+\sum_{\substack{v_i \in \mathcal{I}' \\ v_i \neq v}} \phi(S_{\theta}(u, v_i)-S_{\theta}(u,v))\right\},
\end{equation}
which substitutes $R_{\phi}(u, v; \theta)$ defined in \eqref{crol_alpha_phi} and \eqref{crol_lambda} in our original model. We set $n_{rn}$ and $n_{bs}$ as 10 and 256 for Amazon Books dataset, and as 10 and 512 for the larger Taobao dataset. The number of embedding dimensions and hidden dimensions are set to 32. The margin $m$ of hinge comparison kernel is set to 5. We use the Adam optimizer \cite{adam} with learning rate 0.02 for training.

\noindent {\bfseries Evaluate Metrics.} We use Recall@N shown in \eqref{raw_recall} as the metric to evaluate the performance of the retrieval model. To verify the customization ability of our method for different retrieval ranges, we set $N$ as $50$, $100$, $200$, and $500$, respectively. All the results are percentage numbers with '\%' omitted.

\subsection{Performance Comparison (RQ1)}

\begin{table}[h]
\setlength\tabcolsep{2pt} 
  \caption{Experimental results of comparison between CROLoss and conventional methods. R is short for Recall Metrics.}
  \label{conventional_compare}
  \begin{tabular}{llllll}
    \toprule
    {\bfseries Datasets} & {\bfseries Methods} & {\bfseries R@50} & {\bfseries R@100} & {\bfseries R@200} & {\bfseries R@500}\\
    \midrule
    Amazon & cross-entropy\ loss & \underline{9.68} & \underline{13.24} & \underline{17.46} & 24.26\\
    \  & triplet\ loss & 7.53 & 11.21 & 15.93 & 24.11\\
    \  & BPR\ loss & 8.24 & 12.08 & 16.96 & \underline{25.21}\\
    \cdashline{2-6}
    \  & CROLoss$^{^1}$ & {\bfseries 10.20} & 14.03 & 18.63 & 26.06\\
    \  & CROLoss-lambda$^{^2}$ & 10.17 & {\bfseries 14.07} & {\bfseries 18.81} & {\bfseries 26.20}\\
    \midrule
    Taobao & cross-entropy\ loss & \underline{4.71} & \underline{6.59} & \underline{9.01} & \underline{13.13}\\
    \  & triplet\ loss & 2.46 & 3.71 & 5.43 & 8.84\\
    \  & BPR\ loss & 2.89 & 4.33 & 6.35 & 10.25\\
    \cdashline{2-6}
    \  & CROLoss$^{^3}$ & 4.75 & 6.65 & 9.06 & 13.13\\
    \  & CROLoss-lambda$^{^4}$ & {\bfseries 5.27} & {\bfseries 7.35} & {\bfseries 10.01} & {\bfseries 14.57}\\
    \bottomrule
  \end{tabular}
  \begin{tablenotes}[flushleft]
        \footnotesize
        \item[] 1. Use softplus as kernel and set $\alpha$ to 1.0.
        \item[] 2. Use sigmoid as kernel 1 and softplus as kernel 2 and set $\alpha$ to 1.0.
        \item[] 3. Use exponential as kernel and set $\alpha$ to 1.4.
        \item[] 4. Use sigmoid as kernel 1 and exponential as kernel 2 and set $\alpha$ to 1.4.
    \end{tablenotes}
\end{table}

We compare the best performing CROLoss and CROLoss-lambda with the softmax cross-entropy loss, the triplet loss, and the BPR loss. As Table \ref{conventional_compare} shows, by applying suitable comparison kernel $\phi$ and weighting parameter $\alpha$, our method significantly outperforms these three conventional methods. Furthermore, in almost all cases, the Lambda method helps CROLoss-lambda achieve even better performance. Among the three conventional methods, softmax cross-entropy loss performs the best, while the other two, especially triplet loss, perform not well. We suppose that this is due to the fact that in most practices using triplet loss, the hard triplet mining method is used \cite{defenseTriplet,PinSAGE,fbEmbedding}, while we use the uniform negative sample method in our experiments. We will analyze this phenomenon from the perspective of CROLoss later, and provide an efficient method equivalent to the hard triplet mining to improve their performance.

\subsection{Influence of Kernel and Alpha (RQ2)}

\begin{table*}[t]
\setlength\tabcolsep{2.5pt} 
  \caption{Experimental results of CROLoss with different kernels and weights.}
  \label{crol_main_result}
  \begin{tabular}{p{40 pt}|l|llll|llll|llll|llll}
    \toprule
    {\bfseries Datasets} & {\bfseries Kernels} & \multicolumn{4}{c|}{{\bfseries Recall@50}} & \multicolumn{4}{c|}{{\bfseries Recall@100}} & \multicolumn{4}{c|}{{\bfseries Recall@200}} & \multicolumn{4}{c}{{\bfseries Recall@500}} \\
    \  & \  & $\alpha$=0.6 & $\alpha$=0.8 & $\alpha$=1.0 & $\alpha$=1.2 & $\alpha$=0.6 & $\alpha$=0.8 & $\alpha$=1.0 & $\alpha$=1.2 & $\alpha$=0.6 & $\alpha$=0.8 & $\alpha$=1.0 & $\alpha$=1.2 & $\alpha$=0.6 & $\alpha$=0.8 & $\alpha$=1.0 & $\alpha$=1.2 \\
    \midrule
    Amazon & hinge & 9.03 & 9.50 & \underline{9.88} & 9.53 & 12.95 & 13.35 & \underline{13.55} & 12.87 & 17.64 & \underline{18.04} & 17.94 & 16.76 & 25.56 & \underline{25.63} & 24.99 & 22.91\\
    \  & sigmoid & 9.57 & 9.94 & 9.96 & \underline{10.05} & 13.63 & 13.77 & \underline{13.94} & 13.88 & 18.51 & 18.56 & \underline{18.59} & 18.38 & \underline{{\bfseries 26.59}} & 26.35 & 26.09 & 25.66\\
    \  & exponential & 8.90 & 9.39 & 9.68 & \underline{9.99} & 12.44 & 12.87 & 13.24 & \underline{13.56} & 16.75 & 17.11 & 17.46 & \underline{17.71} & 23.81 & 24.13 & 24.26 & \underline{24.33}\\
    \  & softplus & 9.63 & 9.97 & \underline{{\bfseries 10.20}} & 10.19 & 13.54 & 13.85 & \underline{14.03} & \underline{14.03} & 18.30 & 18.59 & \underline{18.63} & 18.49 & \underline{26.20} & 26.19 & 26.06 & 25.51\\
    \  & lambda$^{^1}$ & 9.65 & 9.80 & 10.17 & \underline{10.19} & 13.58 & 13.78 & \underline{{\bfseries 14.07}} & 14.01 & 18.39 & 18.50 & \underline{{\bfseries 18.81}} & 18.46 & \underline{26.29} & 26.25 & 26.20 & 25.96\\
    \toprule
    \  & \  & $\alpha$=1.0 & $\alpha$=1.2 & $\alpha$=1.4 & $\alpha$=1.6 & $\alpha$=1.0 & $\alpha$=1.2 & $\alpha$=1.4 & $\alpha$=1.6 & $\alpha$=1.0 & $\alpha$=1.2 & $\alpha$=1.4 & $\alpha$=1.6 & $\alpha$=1.0 & $\alpha$=1.2 & $\alpha$=1.4 & $\alpha$=1.6 \\
    \midrule
    Taobao & hinge & 3.75 & 3.96 & \underline{4.02} & 3.72 & 5.37 & \underline{5.62} & \underline{5.62} & 5.09 & 7.51 & \underline{7.82} & 7.68 & 6.84 & 11.35 & \underline{11.64} & 11.18 & 9.63\\
    \  & sigmoid & 3.89 & 4.04 & 4.11 & \underline{4.13} & 5.62 & 5.84 & \underline{5.87} & \underline{5.87} & 7.91 & \underline{8.21} & 8.15 & 8.13 & 11.93 & \underline{12.19} & 12.13 & 11.96\\
    \  & exponential & 4.71 & 4.74 & \underline{4.75} & 4.61 & 6.59 & \underline{6.71} & 6.65 & 6.42 & 9.01 & \underline{9.20} & 9.06 & 8.69 & 13.13 & \underline{13.33} & 13.13 & 12.44\\
    \  & softplus & 4.21 & 4.30 & 4.36 & \underline{4.38} & 5.98 & 6.12 & \underline{6.21} & 6.15 & 8.32 & \underline{8.55} & 8.52 & 8.44 & 12.51 & \underline{12.70} & 12.54 & 12.25\\
    \  & lambda$^{^2}$ & 4.55 & 5.04 & 5.27 & \underline{\bfseries 5.31} & 6.39 & 7.06 & 7.35 & \underline{\bfseries 7.44} & 8.75 & 9.68 & \underline{\bfseries 10.01} & 9.98 & 12.90 & 14.28 & \underline{\bfseries 14.57} & 14.54\\
\bottomrule
  \end{tabular}
  \begin{tablenotes}[flushleft]
        \footnotesize
        \item[] 1. Use sigmoid as $\phi_1$ and softplus as $\phi_2$.
        \item[] 2. Use sigmoid as $\phi_1$ and exponential as $\phi_2$.
    \end{tablenotes}
\end{table*}

We test the CROLoss under different comparison kernels $\phi$ and weighting parameters $\alpha$. The experimental results are shown in Table \ref{crol_main_result}, where the row of lambda kernel represents the performance of CROLoss using the Lambda method under the best choices of $\phi_1$ and $\phi_2$, and will be discussed in detail in the next subsection. For clarity, we underline the best performance of each comparison kernel at each retrieval size $N$, and bold the best performance among all kernels.

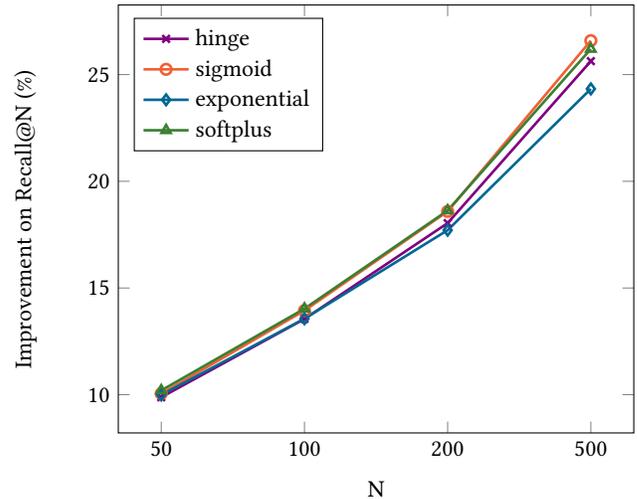
\begin{figure}[h]
\centering
\begin{tikzpicture}
\begin{axis}[
ylabel=Improvement on Recall@N (\%),
xlabel=N,
legend style={
        cells={anchor=west}
      },
legend pos=north west,
symbolic x coords={50,100,200,500},
xtick={50,100,200,500},
    ytick={10,15,20,25},
]
    
\addplot [
    thick,
    color=violet,
    mark=x,
    line width=1pt,
    ] coordinates{
(50, 9.88) 
(100, 13.55) 
(200, 18.04)
(500, 25.63)
}[fill opacity=0.6];

\addplot [
    thick,
    color=RedOrange,
    mark=o,
    line width=1pt,
    ] coordinates{
(50, 10.05) 
(100, 13.94) 
(200, 18.59)
(500, 26.59)
}[fill opacity=0.8];

\addplot [
    thick,
    color=MidnightBlue,
    mark=diamond,
    line width=1pt,
    ] coordinates{
(50, 9.99) 
(100, 13.56) 
(200, 17.71)
(500, 24.33)
}[fill opacity=0.95];

\addplot [
    thick,
    color=OliveGreen,
    mark=triangle,
    line width=1pt,
    ] coordinates{
(50, 10.20) 
(100, 14.03) 
(200, 18.63)
(500, 26.20)
}[fill opacity=0.7];

\legend{hinge, sigmoid, exponential, softplus}
\end{axis}
\end{tikzpicture}

\caption{The Recall@N metrics of four different comparison kernels of CROLoss at different retrieval size $N$ on Amazon dataset.}
\label{fig:kernel_gap}
\end{figure}

By comparing the performance of different comparison kernels in the two datasets, we can see that the performance of the exponential kernel on the Taobao dataset, which contains a larger candidate set, is significantly better than its performance on the Amazon dataset. Figure \ref{fig:kernel_gap} shows a similar phenomenon that the performance of exponential kernel gradually decreases with the increase of $N$ compared to other kernels. Therefore, we suppose that the exponential function is an ideal kernel for the scenario where the ratio of the retrieval size $N$ to the size of the candidate set $|\mathcal{I}|$ is very small. For other comparison kernels with relatively stable gradients, the retrieval size has no significant effect on its performance. Specifically, the softplus kernel is a good choice and can always achieve the best or sub-best performance; the sigmoid kernel performs slightly worse, but it can best match the weighting parameter $\alpha$ to be customized for different retrieval sizes; the hinge kernel is not recommended because it performs poorly and has more hyperparameters. 

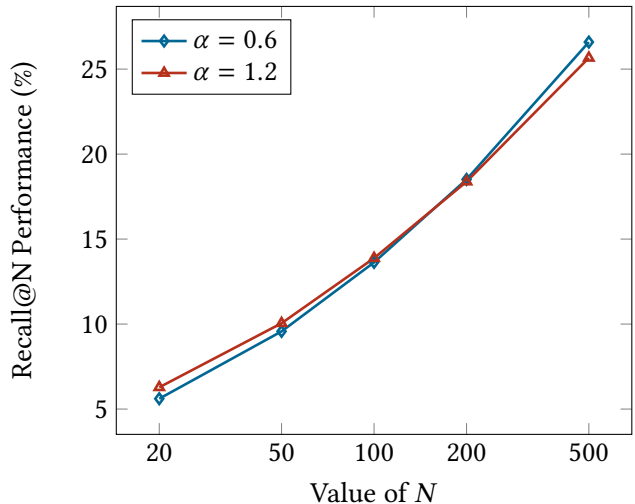
\begin{figure}[h]
\centering
\begin{tikzpicture}
\begin{semilogxaxis}[
	xlabel={Value of $N$},
	ylabel={Recall@N Performance (\%)},
    legend style={at={(0.03,0.97)},
    anchor=north west},
    font=\Large,
    log ticks with fixed point,
    xtick={20,50,100,200,500},
    ytick={5,10,15,20,25,30},
    ]
]

\addplot [
    thick,
    color=MidnightBlue,
    mark=diamond,
    line width=1pt,
    ] coordinates{
(20, 5.61) 
(50, 9.57) 
(100, 13.63) 
(200, 18.51)
(500, 26.59)
}[fill opacity=0.7];

\addplot [
    thick,
    color=BrickRed,
    mark=triangle,
    line width=1pt,
    ] coordinates{
(20, 6.28) 
(50, 10.05) 
(100, 13.88) 
(200, 18.38)
(500, 25.66)
};

\legend{$\alpha=0.6$,$\alpha=1.2$}
\end{semilogxaxis}
\end{tikzpicture}
\caption{The Recall@N metrics of CROLoss with sigmoid kernel and different weighting parameter $\alpha$ on the Amazon dataset.}
\label{fig:alpha_recall}
\end{figure}

By observing the performance under different $\alpha$, we can find that in most cases, a larger $\alpha$ makes CROLoss pay more attention to the top-ranked samples, resulting in better performance of Recall@N under smaller retrieval size $N$. Conversely, a smaller $\alpha$ can achieve better performance with a larger $N$. Among all the comparison kernels, the logistic sigmoid performs best in this customizability, which we suppose is due to it can estimate the $R_{\phi}(u, v; \theta)$ most accurately. Finally, we test CROLoss on a wider range of $N$ using the sigmoid kernel, and the results reported in Figure \ref{fig:alpha_recall} visually demonstrate that CROLoss can be customized for different retrieval sizes by simply adjusting $\alpha$.

\subsection{Performance of Lambda Method (RQ3)}

\begin{table}[h]
\setlength\tabcolsep{1pt}
  \caption{Experimental results of the Lambda method with different kernels. We set $\alpha$ as 1.0 for Amazon dataset and 1.4 for Taobao dataset. R is short for Recall Metrics.}
  \label{lambda_test}
  \begin{tabular}{lllllll}
    \toprule
    {\bfseries Dataset} & {\bfseries Kernel $\phi_1$} & {\bfseries Kernel $\phi_2$} & {\bfseries R@50} & {\bfseries R@100} & {\bfseries R@200} & {\bfseries R@500}\\
    \midrule
    Amazon & unit step & hinge & 9.07 & 12.865 & 17.46 & 24.96\\
    \  & unit step & exponential & 8.74 & 12.15 & 16.35 & 23.20\\
    \  & unit step & softplus & 9.98 & 13.85 & 18.53 & 25.96\\
    \  & sigmoid & hinge & 9.435 & 13.305 & 17.995 & 25.69\\
    \  & sigmoid & exponential & 8.79 & 12.10 & 16.26 & 22.90\\
    \  & sigmoid & softplus & {\bfseries 10.17} & {\bfseries 14.07} & {\bfseries 18.81} & {\bfseries 26.20}\\
    \midrule
    Taobao & unit step & hinge & 3.30 & 4.79 & 6.79 & 10.30\\
    \  & unit step & exponential & 4.85 & 6.81 & 9.36 & 13.70\\
    \  & unit step & softplus & 4.25 & 6.07 & 8.47 & 12.62\\
    \  & sigmoid & hinge & 3.82 & 5.49 & 7.70 & 11.60\\
    \  & sigmoid & exponential & {\bfseries 5.27} & {\bfseries 7.35} & {\bfseries 10.01} & {\bfseries 14.57}\\
    \  & sigmoid & softplus & 4.74 & 6.71 & 9.26 & 13.62\\
    \bottomrule
  \end{tabular}
\end{table}

The performance of CROLoss using the Lambda method has been shown in Tables \ref{conventional_compare} and \ref{crol_main_result}. Compared with the original CROLoss, it has a greater improvement in the Recall@N metric, especially on the Taobao dataset. Moreover, since the selection of $\phi_1$ and $\phi_2$ is independent, CROLoss-lambda can choose a suitable kernel function such as sigmoid to accurately estimate $R_{\phi_1}(u, v; \theta)$, which inherits the good customizability of CROLoss using the sigmoid kernel. Table \ref{lambda_test} shows the impact of different $\phi_1$ and $\phi_2$ choices on CROLoss-lambda, where the choice of $\phi_2$ confirms the previous conclusions: the exponential function is suitable for the Taobao dataset that contains a larger candidate set, and the softplus kernel performs well in the Amazon dataset. For the choice of $\phi_1$, the experimental results show that although the unit step function can calculate $R_{\phi_1}(u, v; \theta)$ more accurately, the smoother logistic sigmoid function can achieve the best performance in most cases.

\subsection{Industrial Results (RQ4)}

We deploy the CROLoss method onto the retrieval stage of our large-scale online advertising system. As mentioned above, the two-tower recommendation model served as the base retrieval model in our system. After completing the model training, we infer and store the representation vectors of all advertised items. When serving online, our model first computes the representation vector of visiting users and then retrieves top-N items from the item vector pool by a fast nearest neighbor searching method. We deployed both the model trained with CROLoss and softmax cross-entropy loss on the A/B test platform. In a fourteen-day online A/B test, our new method improved the Recall metric by 6.50\%, and increased revenue by 4.75\% in Cost-per-click advertising.

\subsection{A New Perspective on Hard Negative Mining}

\begin{figure}[h]
  \centering
\begin{tikzpicture}
\begin{axis}[
x tick label style={
/pgf/number format/1000 sep=},
ylabel=Improvement on Recall@N (\%),
xlabel=N,
enlargelimits=0.2,
legend style={
        cells={anchor=west}
      },
legend pos=north east,
symbolic x coords={50,100,200,500},
ybar=5pt,
bar width=10pt,
nodes near coords,
]
    \addplot[style={black,fill=Tan}]
    coordinates {(50,2.35) (100,2.34)
    (200,2.01) (500,0.88)}[fill opacity=1];
    \addplot[style={black,fill=MidnightBlue}]
    coordinates {(50,1.96) (100,1.95)
    (200,1.67) (500,0.85)}[fill opacity=0.7];
\legend{Triplet Loss,BPR Loss}
\end{axis}
\end{tikzpicture}
  \caption{Performance improvement of using CROLoss for hard negative mining on triplet loss and BPR loss.}
  \label{fig:triplet_bpr_alpha}
\end{figure}
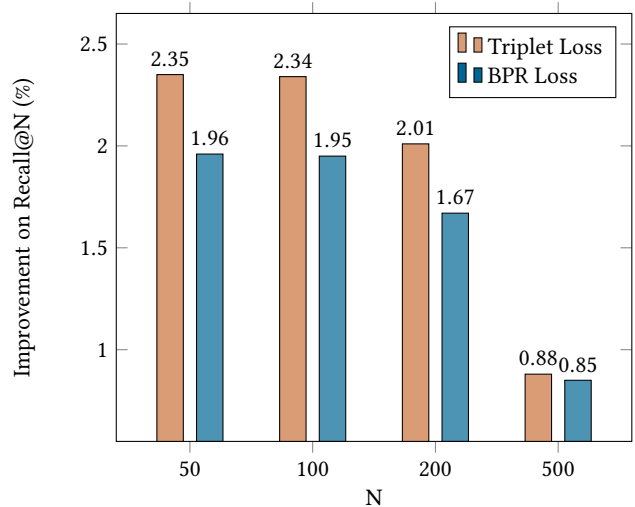


As shown in \eqref{crol_sfx}, \eqref{crol_tri}, and \eqref{crol_bpr}, the softmax cross-entropy loss, the triplet loss, and the BPR loss can be regarded as special cases of CROLoss. This provides a new perspective to understanding why hard negative mining (triplet mining) \cite{triplet, fbEmbedding} is more commonly used in the practice of triplet loss than in cross-entropy loss. In addition to different kernels, the difference between triplet loss and cross-entropy loss is that the former is equivalent to setting $\alpha=0$ and the latter's $\alpha=1$. BPR loss can also be classified into the category with $\alpha=0$. 

If we regard the weighting function $w_\alpha(x)$ as the probability density function of an importance sampling, when $\alpha=0$, it is equivalent to uniform sampling, and when $\alpha>0$, it is equivalent to paying more attention to the top-ranked samples of $(u,v)$. For these $(u,v)$, most of the easy negative samples are already well optimized, so the gradient is mostly brought by the hard negative samples. Therefore, the CROLoss with $\alpha>0$ can be seen as an efficient hard negative mining method for hard negatives, which interprets why cross-entropy does not require additional work on hard negative mining but triplet loss does. Figure \ref{fig:triplet_bpr_alpha} shows the performance improvement over the original loss by using the same kernel as triplet loss and BPR loss in CROLoss and setting $\alpha=1$. Specifically, this effective mining method can significantly improve the performance of both loss functions, especially when N is smaller.

\section{Conclusion}
This paper proposes the novel Customizable Recall@N Optimization Loss, an objective direct optimize the Recall@N metrics and can customized for the retrieval size $N$ required by different applications. We achieve the goal by rewriting the definition of Recall@N via a ranking statistic $R_\psi(u, v; \theta)$ and assigning an importance weight for each Recall@N metric via the weighting function $w_\alpha$. To optimize the CROLoss, we introduce the comparison kernel function $\phi$ that makes the CROLoss differentiable. Furthermore, we develop the Lambda method that solves the CROLoss more efficiently. We test our method on two real-world public datasets and our online advertising system, and the result shows our method outperforms the conventional objectives on Recall@N metrics and proves the ability to customize $N$ for the Recall@N optimization problem.


\bibliographystyle{ACM-Reference-Format}
\balance
\bibliography{sample-sigconf}

\end{document}